\documentclass[twocolumn,pra]{revtex4}

\usepackage{bm,ulem,graphicx,amsbsy,amsbsy,amsmath,amsfonts,amssymb,amsthm}
\usepackage{color,mathrsfs,latexsym,dsfont}

\newcommand{\iea}[0]{{\it et al.~}}

\newcommand{\eeqref}[1]{Eq.~(\ref{#1})}

\definecolor{MyGreen}{rgb}{0.0,0.5,0.1}
\def\ket#1{|\,#1\,\rangle}

\begin{document}

\title{Optical quantum memory}
\author{A. I. Lvovsky$^{1}$, B. C. Sanders$^{1}$, W. Tittel$^{1}$}
\affiliation{Institute for Quantum Information Science, University of Calgary, Alberta, Canada T2N 1N4}


\begin{abstract}
Quantum memory is important to quantum information processing in many ways:
a synchronization device to match various processes within a quantum computer,
an identity quantum gate that leaves any state unchanged,
and a tool to convert heralded photons to photons-on-demand.
In addition to quantum computing, quantum memory would be instrumental for the implementation of long-distance quantum communication using quantum repeaters.
The importance of this basic quantum gate is exemplified by the multitude of
optical quantum memory mechanisms being studied: optical delay lines, cavities, electromagnetically-induced transparency,
photon-echo, and off-resonant Faraday interaction.
Here we report on the state-of-the-art in the field of optical quantum memory,
including criteria for successful quantum memory and current performance levels.
\end{abstract}

\maketitle

\section{Introduction}
\subsection{Background}
\label{sec:background}

Quantum information science incorporates quantum principles into information processing and communication.
Amongst the most spectacular discoveries and conjectures, we know that quantum cryptography could enable information-theoretic secure communication through public channels \cite{GRTZ02}, and
quantum computing would efficiently solve certain computational problems that are
believed to be intractable by conventional computing \cite{Mer07}.
Furthermore quantum dynamics becomes efficiently simulatable on a quantum computer \cite{Lloyd96}.
The prototypical model of quantum information processing represents information as strings of qubits, and processing is effected by unitary quantum gates.

The qubit is a single-particle state in a two-dimensional Hilbert space.
If the particle is a single photon, then the qubit can be encoded in several ways.
For example, in polarization encoding
the logical zero state~$\left|0\right\rangle$ can correspond to a single photon
being left-circularly polarized and~$\left|1\right\rangle$ to
right-circularly polarized.
Other examples include path  \cite{KLM01},
photon-number \cite{LR02,BLS06},
and time-bin encodings \cite{MdRT+02}.
A general qubit state can be expressed as a superposition of
$\left|0\right\rangle$ and~$\left|1\right\rangle$,
and general states of quantum information are superpositions of strings of qubits.
Quantum memory needs to store qubit strings or parts thereof faithfully and
to release them on demand.

Storage of a quantum state need not be perfect.
Fault-tolerant quantum error correction can be employed to make
an imperfect memory sufficient as long as the fidelity of the memory `gate'
exceeds a particular performance threshold \cite{Sho95,Got06}.
Next we study the specific requirements for optical quantum memory to be effective
for quantum information tasks.

\subsection{Performance criteria}
\label{sec:requirements}

In general quantum memory stores a pure or mixed state represented by density matrix~$\rho$
and outputs a state~$\rho'$, which should be close to $\rho$.
The ultimate performance criterion for quantum memory is
the {\it worst-case fidelity} with respect to the set of input states
where the fidelity for a specific state is given by
$F(\rho)=\text{Tr}\sqrt{\sqrt{\rho'}\rho\sqrt{\rho'}}$.
There exists a threshold worst-case fidelity beyond which fault-tolerant quantum
error correction methods can overcome memory imperfection \cite{Sho95,Got06}.

The fidelity of quantum-optical memory for an arbitrary set of input states can be determined by subjecting it to complete quantum process tomography \cite{Lobino08,Lobino09}. However, this procedure is relatively bulky, so in practical experimental implementations, other performance criteria are used.
For example the term `fidelity' sometimes refers to state overlap $\left\langle\psi\left|\rho'\right| \psi\right\rangle$ (possibly after post-selection), which is the square of $F(\rho)$
for the case that $\rho=\left|\psi\rangle\langle\psi\right|$ is a pure state.
{\it Average fidelity} is often used, where the average is taken over all
input states with respect to an assumed prior distribution \cite{HPC05}.

Another popular criterion is {\it efficiency}~$\eta$,
which is the ratio between the energies of the stored and retrieved pulses.
Efficiency,
while easy to determine experimentally, does not account for possible detrimental effects such as contamination of the retrieved state by the excess noise from the storage medium.

In continuous-variable implementations, memory can be characterized by the {\it transfer coefficient} and {\it conditional variance} \cite{HPJ+08}.
These quantities are convenient for characterizing quantum memory for single-mode fields
provided that input and retrieved states are Gaussian.

The {\it multimode capacity} of quantum memory determines the number of optical modes that can be stored in the memory cell with the requisite performance threshold or better.
The multimode capacity strongly depends on the memory mechanism \cite{NRL+08,ASdRG09}.

Quantum memory needs to be able to store the state long enough to perform the task at hand
so {\it storage time} is another essential memory performance criterion. For many applications, an appropriate figure of merit would be the \textit{delay-bandwidth product}, i.e the ratio between storage time and duration of the stored pulse.

\subsection{Applications}

In {\it optical quantum computation} the role of quantum memory is to store  quantum bits so that operations can be timed appropriately. Many qubits are being processed in parallel with each other at each step in time,
and these processing steps must be synchronized  \cite{KLM01,Raussendorf2001,Kok2007}.

{\it Quantum communication} suffers from imperfect transmission channels, resulting, for example, in quantum key distribution being possible only over finite distances.
The {\it quantum repeater} \cite{BDCZ98,Sangouard2009} solves this problem and permits quantum communication over arbitrary distances with a polynomial cost function. A necessary component of the quantum repeater is quantum memory, which, similar to quantum computation, allows synchronization between entangled resources distributed over adjacent sections of the transmission link.

In addition, quantum memory for light finds applications in {\it precision measurements} based on quantum interference of atomic ensembles. By transferring quantum properties of an optical state to the atoms one can reduce the quantum noise level of the observable measured, thereby improving the precision of magnetometry, clocks, and spectroscopy  \cite{Appel09}.

Finally, quantum optical memory can be used as a component
of {\it single-photon sources}. If a single-photon detector is placed in
one of the emission channels of non-degenerate spontaneous parametric
down-conversion, a detection event indicates emission of a
photon pair, and thus the presence of a single photon in the other
channel  \cite{Hong1986,GRA86}. Such a heralded photon is emitted at an arbitrary time,

In the following, we review recent theoretical and experimental work related to different approaches to quantum memory.

\section{Realizations}

\subsection{Optical delay lines and cavities}
The simplest approach to storage of light is an optical delay line, e.g. an optical fibre. This approach has been used to synchronize photons with the occurrence of certain events \cite{LvHBZG07}.
However, the storage half-time, i.e.\ the time after which half the photons are lost, is, in the case of 1.5 $\mu$m wavelength and telecommunication fibres, limited to around 70$\mu$s, corresponding to a fibre length of $\sim$15 km. The half-time decreases at other wavelengths due to increased loss. Furthermore, the storage time in an optical delay is fixed once a delay length is chosen, contrary to the requirement of variable, on-demand output as desired in most applications.

Alternatively, light can be stored in a high-Q cavity. The light effectively cycles back and forth between the reflecting boundaries, and can be injected into and retrieved from the cavity  using electro-optical or non-linear optical means \cite{PF02a,PF02b,LR06}, or by quantum state transfer with passing atoms \cite{MHN+97}. For example, light has been stored for over a nanosecond in wavelength-scale photonic crystal cavities with a tunable $Q$ factor which allows control over the storage time \cite{TNK+06}. Dynamic control of $Q$ can be achieved by adiabatically tuning the frequency of the stored light, which has yielded output pulses as short as $\sim 0.06$ns, much shorter than the storage time \cite{TNTK09}.

Unfortunately, storage of light in cavities suffers from the trade-off between a short cycle time and a long storage time, limiting the efficiency or the delay-bandwidth product. Therefore, whereas optical delay lines and nanocavities could be appropriate for obtaining on-demand single photons from heralded sources \cite{Hong1986,GRA86}, they may not be suitable for quantum memory or quantum repeaters.

\subsection{Electromagnetically-induced transparency}
\label{subsec:EIT}
\subsubsection{Background}

Electromagnetically-induced transparency (EIT) is a
nonlinear optical phenomenon observed in atoms with
an energy-level structure resembling the letter~$\Lambda$ [Fig.~1(a)].
Two optical fields couple the excited level to their respective
ground levels: the weak signal field which may carry the
quantum information load and the strong control field
which is used to steer the atomic system.

If the control field is absent, the signal field, which interacts with the
resonant two-level system, undergoes partial or complete absorption.
In the presence of the control field, the
absorption of the signal is greatly reduced whenever the
frequency difference of the two optical fields is close to the
frequency of the Raman transition between the ground
states of the~$\Lambda$ system (the condition known as the two
photon resonance) [Fig.~1(b)].

Transparency is observed
when the fields are detuned from the two-photon resonance by no more than
\begin{equation}
\label{WEIT}
	W_{\rm EIT}\sim\frac 4 {\sqrt{\alpha L}}\frac{\Omega^2}{W_{\rm line}}
\end{equation}
for~$\alpha$ the absorption coefficient of the medium in the absence of EIT,
$L$ its length, $\Omega$ the Rabi frequency of the control field and~$\hbar W_{\rm line}$ the uncertainty of the excited level energy associated with homogeneous or inhomogeneous broadening.
The width of the EIT window is thus proportional to the intensity of the control field and can be much narrower than the $W_{\rm line}$.

\begin{figure*}
\includegraphics[width=\textwidth]{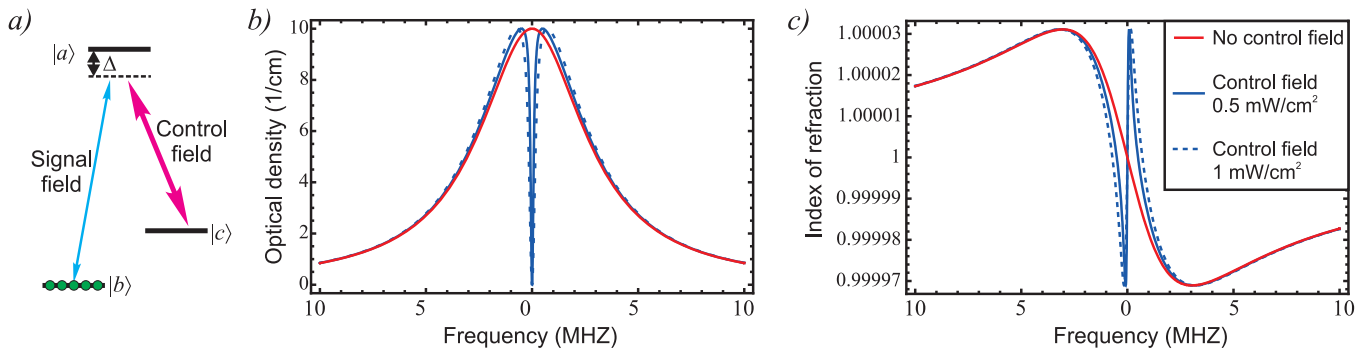}
\caption{
    Electromagnetically-induced transparency. (a) Atomic level configuration.
    Both fields are detuned from the resonance by the same frequency so the two-photon resonance condition is fulfilled.
    (b)~Optical density,
    (c)~index of refraction of an ensemble of atoms in the absence (red) and in the presence (blue) of EIT.
    In spite of a significant optical depth, the variation of the index of refraction is very small.
    The atomic parameters used to generate the plots correspond to a cloud of ultracold rubidium atoms.  }
 \label{EITConcept}
\end{figure*}

EIT is largely insensitive to the detuning $\Delta$ of the optical fields from their respective individual transitions. It can thus exist in the presence of inhomogeneous broadening as long as both optical transitions experience the same frequency shift.
A good example is provided by atoms in a warm gas whose transitions are broadened by the Doppler effect associated with atomic motion.
If the ground levels are of similar energy and the two optical fields are co-propagating, the Doppler shifts cancel and EIT is observed. Counterintuitively, according to Eq.~(\ref{WEIT}), EIT window in the presence of inhomogeneous broadening is much narrower than in the case of pure homogeneous broadening \cite{JKLS02}.

EIT was first demonstrated in 1991 for strontium vapor \cite{BIH91}, and since has been observed in various media.
Among the most popular physical systems, particularly in application to quantum memory, are ensembles of alkali atoms and rare-earth doped solids.

\subsubsection{Slow light and memory}
\label{subsubsec:qmemoryEIT}


According to the Kramers-Kronig relations, an anomaly in the absorption spectrum always comes together with an anomaly in dispersion.
A light field propagating under EIT conditions experiences large positive dispersion [Fig.~\ref{EITConcept}(c)],
which implies reduction of the group velocity by a factor of
$n_g\sim \alpha c W_{\rm line}/4\Omega^2$
with respect to the velocity of light in vacuum $c$ \cite{KJYH95,FL02}.
The group velocity can in theory be arbitrarily reduced by lowering the intensity of the control field. Experimentally, slowdown by up to seven orders of magnitude have been observed \cite{HHDB99,BKRY99}.

\begin{figure}
\includegraphics{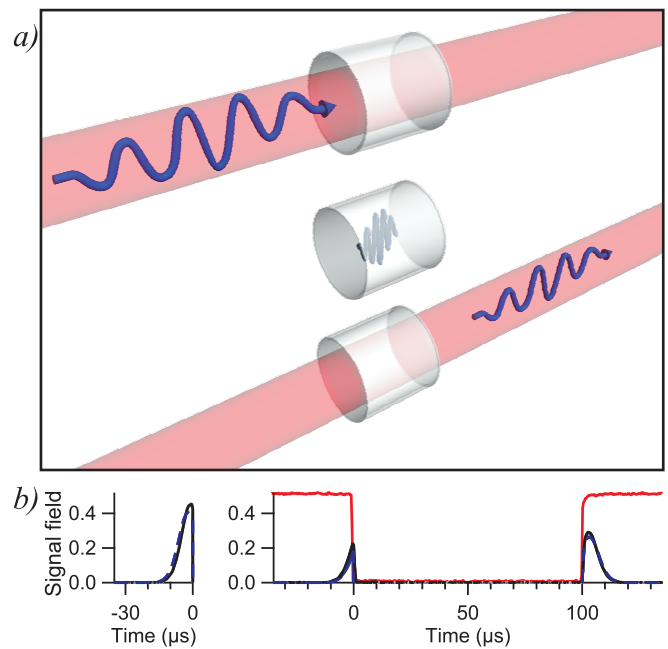}
 \caption{Storage of light by means of electromagnetically-induced transparency. (a) Idealized picture. The signal pulse enters the cell under the EIT conditions (with control field on, top image). While the spatially compressed pulse propagates inside the EIT cell, the control field is switched off, so the quantum information carried by the pulse is stored as a collective excitation of the ground states (middle image). When the pulse needs to be retrieved, the control field is switched back on (bottom image). (b)~Optimized classical light storage in a rubidium vapor cell with a buffer gas, $\alpha L=24$. The red curve shows the control field, solid black --- experimental signal, dashed blue --- theoretical signal. Left: input signal pulse of optimal shape. Right: storage and retrieval. Because of the finite optical depth, the signal pulse does not entirely fit into the cell, resulting in a fraction of the pulse leaking through the cell before the control field is turned off. From Fig.~1(d) of Ref.~ \cite{PNG08}.
 }
 \label{EITStorage}
\end{figure}

Slow light is a `trademark' property of EIT and has a variety of applications, for example buffering of optical communication traffic \cite{BBB08}.
It is also the basis for the quantum memory application,
which functions as follows [Fig.~\ref{EITStorage}(a)]. A light pulse, resonant with the EIT window, enters the EIT medium and slows down.
The slowdown entails spatial compression, so the pulse, whose initial spatial extent by far exceeds $L$, will fit inside the medium. Once it is inside, we adiabatically reduce the control field intensity and bring the group velocity down to zero, thereby `collapsing' the EIT window and storing the pulse in the medium. When the pulse needs to be retrieved, the control field is turned back on. The pulse then resumes its propagation and leaves the EIT medium.

Before the signal has entered the EIT medium, all atoms are optically pumped by the control field into the ground level $\left|b\right\rangle$ so the initial atomic state is
\begin{equation}
\label{eq:psi0}
    \left|\psi_0\right\rangle=\left|b_1\dots b_N\right\rangle.
\end{equation}
After the signal pulse has entered the medium and been stored, its quantum state is transferred into a {\it collective excitation} of the atoms in the EIT medium. For example, if the signal state is a single photon, the atomic state becomes (neglecting normalization)
\begin{equation}\label{eq:psi1}
    \left|\psi_1\right\rangle_A=\sum_j\psi_je^{i\Delta kz_j}\left|b_1\dots c_j \dots b_N\right\rangle.
\end{equation}
for~$N$ the number of atoms in the ensemble, $z_j$ the position of atom $j$ along the field propagation and $\Delta k$ the difference in the wavevectors of the control and signal fields.
In other words, one of the atoms is transferred into the other ground state, $\left|c\right\rangle$, but it is not known which atom it is.

In the ideal case of absent ground state decoherence, neither during the transfer, nor during the storage period does the atomic state contain any fraction of the excited state $\left|a\right\rangle$. This means that the EIT-based memory is not affected by spontaneous decay of that state.

The concept of EIT-based quantum memory has been put forward by Fleischhauer and Lukin in 2000~ \cite{FL00} and described in detail in Refs.~ \cite{FL02,Luk03,FIM05}. First experimental demonstrations have been reported in 2001~ \cite{PFMW01,Hau01}. In Ref.~ \cite{PFMW01} 10--30 $\mu$s pulse at a wavelength of 795 nm has been stored in a 4-cm rubidium vapor cell for up to 0.2 ms. In the work~ \cite{Hau01}, a magnetically trapped thermal cloud of sodium atoms has been used and the memory decay time of 0.9 ms was observed.  Verification of memory performance consisted of measuring the intensity of the stored and retrieved pulses.

Gorshkov  et al. \cite{GAF+07} developed a detailed theory of EIT-based storage of light for a variety of experimental configurations, providing techniques for optimization of the memory performance. In the case of optimal matching of the temporal shapes of the input signal and control fields, optical depth  $\alpha L$ of the storage medium outside the EIT window is the only parameter that determines the storage efficiency. For efficient storage, $\alpha L$ must significantly exceed one.

This requirement can be understood as follows. First, the spectrum of the signal pulse must fit within the transparency window. This implies that the signal pulse duration must satisfy $\tau\gg 1/W_{\rm EIT}$, where $W_{\rm EIT}$ is given by Eq.~(\ref{WEIT}).
Second, the pulse must fit geometrically within the EIT medium: the spatial extent of the signal pulse, compressed due to the slow light effect and given by $c\tau/n_g\gg\sqrt{L/\alpha}$, must not exceed $L$.
These bounds can be satisfied at the same time if the slowdown is sufficient, which translates into a demand for a high contrast of the EIT window, i.e. high $\alpha L$.

Achieving high optical density is a challenge in many optical arrangements including magneto-optical traps and solid state systems. In vapor cells, higher atomic density will increase $\alpha L$, but also will degrade the EIT due to increased ground state decoherence and competing processes such as four-wave mixing and stimulated Raman scattering \cite{PNG08}.

The findings of Gorshkov {\it et al.} were verified in an experiment in warm rubidium vapor. For moderate optical densities ($\alpha L \lesssim 25$), the experimental results showed excellent agreement with a three-level theoretical model without any free parameters [Fig.~2(b)] \cite{NGP+07,NPG08,PNG08}. For higher optical densities, four-wave mixing effects come into play, leading to an additional (idler) optical mode being generated. Although the associated parametric gain may lead to better compression of the pulse \cite{CVH09}, it also brings about additional quantum noise that degrades the storage fidelity.

EIT-based memory can be implemented in solid media, with the advantage of significantly longer storage times (see Sec. \ref{sec:decoherence}).  Following an initial observations of EIT \cite{Ichimura98} as well as ultraslow and stored light \cite{TSS+02} in praseodymium doped Y$_2$SiO$_5$ crystal, Longdell  et al.\ stored light in a similar crystal with a decay time of 2.3 seconds \cite{LFSM05}. A disadvantage of this crystal in application to light storage is a relatively low optical density, which is due to the inhomogeneous broadening associated with the difference in ionic radii of Y$^{3+}$ and Pr$^{3+}$. Attempts to increase the dopant concentration only result in a broader line without increasing the optical density. Recently, EIT has been observed in another praseodymium doped crystal, La$_2$(WO$_4$)$_3$, which exhibits an inhomogeneous broadening that is 15 times smaller than Pr$^{3+}$:Y$_2$SiO$_5$ but at the cost of significantly increased homogeneous decay \cite{GG-NB+09}.

We now review a few experiments in which quantum states of light have been stored, and the retrieved pulses were demonstrated to retain some of the nonclassical properties. In 2005, single photons, generated using the DLCZ method (see below), have been stored a cold atom cloud \cite{CMJ+06} and in a vapor cell \cite{EAM+06} and retrieved about 0.5$\mu$s later. Sub-Poissonian statistics of the retrieved light have been verified using a Hanbury Brown \& Twiss detection scheme.

In an important step towards applications, a dual-rail single-photon qubit has been stored in 2008 \cite{CDLK08}.
A single photon has been split into two spatially separate optical modes, each of which has been stored in a cloud of ultracold optical atoms. Upon retrieval, the modes were recombined and subjected to an interference measurement, which demonstrated that not only nonclassical photon statistics, but also the phase relation between the two stored modes has been preserved. This experiment constitutes the first mapping of an optical entangled entity in and out of quantum memory.

In 2004, propagation of another quantum information primitive, squeezed vacuum,
has been observed under EIT conditions \cite{Kozuma04}, which followed by experimental demonstrations of storage of squeezed vacuum in 2008 \cite{HAA+08,Honda09,AFK+08}. Quadrature noise of the retrieved pulses was measured by means of homodyne detection and observed to be below the shot noise level, demonstrating that some of the initial squeezing has been preserved through the storage procedure.

EIT-based storage of quantum light suffers from background
noise in the retrieved signal. This noise is likely to originate from the  repopulation of the state $\left|c\right\rangle$ \cite{FVAL06} associated, for example, with the atomic drift into and out of the interaction area. In the presence of the control field, the atoms are pumped from $\left|c\right\rangle$ into the excited state $\left|a\right\rangle$, and then spontaneously decay back into the ground state emitting thermal photons that contaminate the signal mode. This effect is negligible when a macroscopic pulse is stored and its classical properties (such as the energy and the pulse shape) are measured upon retrieval. On the other hand, the detrimental effects of the noise become significant when the quantum properties of the storage process are of interest. To minimize this noise, most experiments on non-classical light storage had to compromise on storage efficiency and lifetime.

The background noise has been observed in a homodyne detection setting \cite{HHG+06}
and has been investigated theoretically \cite{PJB+05,PJB+05e,HPJ+08}.
These predictions were successfully applied to an experiment on propagation of squeezed light under EIT conditions \cite{FLK+09}.
Unlike the classical case, however, there does not yet exist a comprehensive study where a full quantum-theoretical description of EIT-based light storage would be developed and verified experimentally.

\subsection{The DLCZ Protocol} \label{sec:DLCZ}
\begin{figure}
\includegraphics[width=\columnwidth]{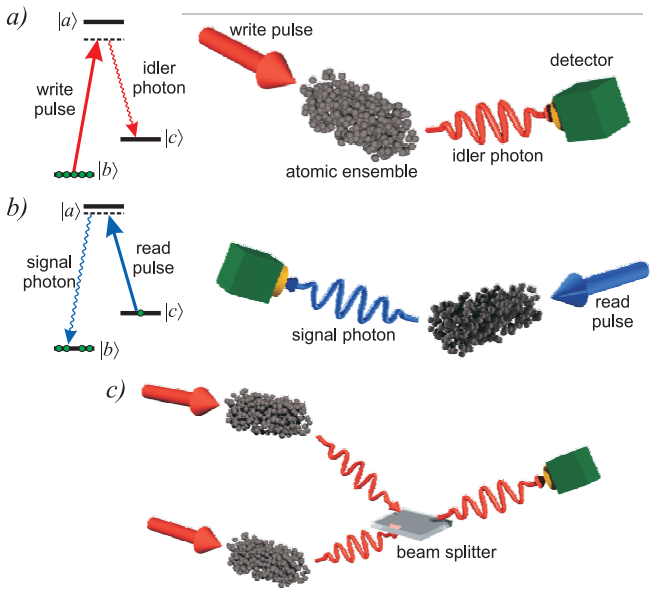}
 \caption{The Duan-Lukin-Cirac-Zoller protocol.
   (a) the atoms, initially in the state $\left|b\right\rangle$, are illuminated with a weak off-resonant {\it write} pulse, resulting in a probability of Raman transfer into state $\left|c\right\rangle$. Each such transfer is associated with scattering of a photon in an arbitrary direction. A single spatial mode of the scattered ({\it idler}) light is selected, e.g. by means of an optical fiber, and subjected to measurement with a single-photon detector. The parameters of the write pulse are chosen so that the probability of a detection event is low. If such an event does occur, it indicates with high likelihood that exactly one photon has been emitted by the atomic ensemble into the detection mode. Spatial filtering erases the information about the location of the atom that has emitted the photon. As a result, the detection event projects the atoms onto a collective superposition of the type \eqref{eq:psi1}.
   (b) Application of a classical {\it read} pulse converts this collective excitation to optical form, resulting in emission of the {\it signal} photon and transfer of the atoms back into the state \eqref{eq:psi0}. (c) Entanglement between two ensembles is created when the detection modes from two samples are overlapped on a beam splitter, so the source of the idler photon becomes indistinguishable.
 }
 \label{DLCZfig}
\end{figure}

Closely related to EIT-based quantum memory is a scheme for creating long-lived, long-distance entanglement between atomic ensembles proposed by Duan, Lukin, Cirac and Zoller (DLCZ) \cite{DLCZ01}. An elementary step of the DLCZ procedure consists of creating a stored collective excitation of the type \eqref{eq:psi1} in an ensemble of $\Lambda$-type atoms. In contrast to regular optical memory, this excitation is produced not by an external photon entering an ensemble, but by the ensemble itself. The atoms, initially in the state $\left|b\right\rangle$, are illuminated with a weak off-resonant optical pulse (called the \textit{write} pulse), resulting in a probability of Raman transfer of atoms into state $\left|c\right\rangle$ (Fig.~\ref{DLCZfig}a). Each such transfer is associated with scattering of a photon in an arbitrary direction.

A single spatial mode of the scattered (\textit{idler}) light is selected, e.g. by means of an optical fiber, and subjected to measurement with a single-photon detector. The parameters of the write pulse are chosen so that the probability of a detection event is low. If such an event does occur, it indicates with high probability that exactly one photon has been emitted by the atomic ensemble into the detection mode.

Spatial filtering erases the information about the location of the atom that has emitted the photon. As a result, the detection event projects the atoms onto a collective superposition of the type \eqref{eq:psi1}. The state of the atomic ensemble becomes equivalent to that as if a single photon has been stored therein using the EIT technique. Therefore one can apply a classical (\textit{read}) field on the $\left|c\right\rangle$-$\left|a\right\rangle$ transition, which will play the role of the EIT control field, leading to retrieval of a \textit{signal} photon from the ensemble and transfer of the atoms back into the state \eqref{eq:psi0} (Fig.~\ref{DLCZfig}(b)).

In its many aspects, the scheme resembles heralded preparation of a single photon from a biphoton generated via parametric down-conversion \cite{Hong1986,GRA86}. A fundamental difference is that the heralded atomic excitation is long-lived and can be retrieved at an arbitrary time, which enables application in a quantum repeater. The protocol can also be viewed as a deterministic single-photon `pistol': once it is `loaded' with an idler detection event, it can `shoot' the signal photon on demand.

Figure~\ref{DLCZfig}c illustrates preparation of a single link of long-distance entanglement between two remote atomic ensembles. The ensembles are simultaneously illuminated with write pulses and the spatial modes in which the idler photon is to be detected are mixed on a beam splitter. Now, if a photon has been detected in one of the beam splitter outputs, it is impossible to tell which of the two ensembles has emitted the photon. As a result, the state of the two ensembles becomes an entangled superposition
\begin{equation}\label{DLCZEnt}
\Psi=\frac 1 {\sqrt 2}\left|\psi_0\right\rangle\left|\psi_1\right\rangle+e^{i\phi}\left|\psi_1\right\rangle\left|\psi_0\right\rangle),
\end{equation}
where the phase $\phi$ depends on the lengths of the optical links between the ensembles and the detection apparatus.

The DLCZ protocol does not constitute quantum memory for light in the strict sense: being stored is not an external, arbitrary state of light but a heralded excitation. Nevertheless, the scheme fully replaces the ``orthodox'' memory as far as the quantum repeater application is concerned. Furthermore, it is more convenient in that it requires no additional nonclassical light sources: the role of these is played directly by the memory cells.

There exists a vast body of experimental work on the DLCZ protocol. The first implementations were reported in 2003 by two groups. Kuzmich et al.\ worked with a cesium MOT and observed nonclassically correlated idler and signal photons for storage times of about 400 ns~ \cite{Kuzmich03}. Van der Wal  et al.\ observed nonclassical correlations between the idler and signal light intensities for a read and write pulse separated by a few hundred nanoseconds~ \cite{vanderWal03}. Regular photodiodes were used rather than photon counters, and both generated pulses were macroscopic.

These initial observations were followed by extensive research aimed at refining and characterizing the scheme~ \cite{Chou04,Eisaman04,Jiang04,Polyakov04,Laurat06}. A particularly significant improvement was achieved thanks to a noncollinear beam geometry in which the signal and idler photons are emitted against a dark background, and can thus be detected without additional filtering~ \cite{Braje04,Balic05}. The conditional probability of generating the signal photon on observation of the idler photon reached a value of 50\%, with a suppression of the two-photon component below 1\% of the value for a coherent state~ \cite{Laurat06}. Most recently, the single photon produced with the DLCZ technique was used to perform quantum tomography on a W state~ \cite{Kimble09}.

The DLCZ protocol has been further enhanced by a feedback procedure \cite{Matsukevich06a,deRiedmatten06,Chen06} in which the write pulses are repeated until an idler photon has been registered by the detector. Upon a detection event, the read pulse is applied at a desired moment in time. In this manner, a good approximation of a deterministic single-photon source can be constructed, with typical unconditional quantum efficiencies on a scale of 10\%.

Enclosing the atomic ensemble into an optical cavity allows increasing its effective optical depth~ \cite{Vuletic05,Vuletic06,Vuletic07}, leading to intrinsic photon retrieval efficiencies of up to 84\%~ \cite{Vuletic07}. An additional advantage of the cavity is that the temporal modes of the emitted signal and idler photons are determined by the cavity parameters, and are thus identical, as can be demonstrated, for example, through the Hong-Ou-Mandel effect~ \cite{Vuletic06}. On the other hand, this scheme introduces additional losses associated with coupling the photons out of the cavity.

Signal photons prepared using the DLCZ method are transform limited. Therefore, the photons emitted by two similarly prepared DLCZ samples are largely indistinguishable. This was demonstrated by observing Hong-Ou-Mandel interference between these photons~ \cite{Felinto06,Chaneliere07,Yuan07}.

The first DLCZ link between atomic ensembles has been demonstrated in 2004 by Matsukevich and Kuzmich~ \cite{Matsukevich04}. Two cylindrical areas of the same atomic cloud were simultaneously excited by a write pulse, and the two corresponding idler modes were mixed on a polarizing beam splitter, so the source of the idler photon becomes indistinguishable. The polarization state of this photon was measured, projecting the ensembles onto the state \eqref{DLCZEnt}. A subsequent pair of read pulses was followed by a polarization measurement of the signal photon, which exhibited Bell-type correlations with the idler. This work was criticized by van Enk and Kimble~\cite{Kimblecomment,Kuzmichreply} for the postselected character of the measurement. An experiment in a similar setting proving the presence of entanglement without resorting to postselection was later reported by the same group \cite{Chou05}.

After propagating through the beam splitter (Fig.~\ref{DLCZfig}c), the idler photon is in an entangled state with the collective excitations in both atomic samples. This entanglement was used in 2008 by Chen  et al.\ to implement quantum teleportation~ \cite{Chen2008}. A Bell-state measurement was performed on the idler photon and a coherent state of arbitrary polarization, teleporting the polarization state of the coherent state onto the atomic excitations. The excitations were then converted into optical form in order to measure the teleportation fidelity.

On application of the write pulse, the idler photon is emitted by a \textit{single} DLCZ sample in an arbitrary direction and, generally, with an arbitrary polarization. This feature was utilized by a number of groups to demonstrate entanglement between various degrees of freedom of the optical and atomic excitations: polarization~ \cite{Matsukevich05}, angular momentum~ \cite{Inoue06}, and direction~ \cite{Chen07}. By the same principle, entanglement of a frequency-encoded optical qubit with a cold mixture of $^{85}$Rb and $^{87}$Rb isotopes has been demonstrated~ \cite{Lan07}.

A photon entangled with the atomic ensemble can then be stored in an EIT-based memory cell~ \cite{Matsukevich06b}, leading to entanglement of two remote atomic qubits. Such entanglement can also be produced by a Bell measurement on a pair of polarization encoded optical qubits obtained from two remote samples~ \cite{Zhao07,Yuan08}. The advantage of this approach in comparison with the classic DLCZ protocol is that the polarization encoded qubits are much less sensitive to fluctuations of optical phases in communication links.

Entanglement swapping between two DLCZ links has been demonstrated by Chou  et al.\ in 2007~ \cite{Chou07}. After preparing entanglement in two pairs of nodes, simultaneous read pulses were applied to two neighboring nodes in different pairs. The generated signal modes were mixed on a beam splitter and subjected to a photon number measurements. Detection of a single signal photon projects the two remaining nodes onto an entangled state, which has been verified by reading out the signal photon from these nodes.


The DLCZ scheme is well suited for quantum repeater applications, but does not directly enable storage of arbitrary quantum information from outside the system. However, the scheme could be used as quantum memory for arbitrary qubits by means of quantum teleportation \cite{Chen2008}. As the entanglement is generated by spontaneous emission, such an optical quantum memory is only useful in a post-selected way.


\subsection{Photon-echo quantum memory}

\subsubsection{Background}
Similar to EIT-based storage discussed in sec.~\ref{subsec:EIT}, photon-echo quantum memory relies on the transfer of the quantum state carried by an optical pulse into collective atomic excitation. However, in contrast to EIT, it takes advantage of the inhomogeneous broadening. Consider again an ensemble of two-level atoms in state (\ref{eq:psi0}). After absorption of a signal photon at $t=0$, the state is given (in un-normalized form) by \cite{ASdRG09}
\begin{equation}
     \left|\psi_1\right\rangle=\sum_j\psi_je^{-i\delta_jt}e^{ikz_j}\left|b_1\dots a_j \dots b_N\right\rangle,
\label{collective state for IB media II}
\end{equation}
where $k$ denotes the wave vector of the signal field,
$\delta_j$ is the detuning of the transition of atom $j$ with respect to the light carrier frequency, and all other variables are as in Eq. \ref{eq:psi1}. Although the atomic dipoles are initially phase aligned with $k$, this alignment rapidly decays because $\delta_j$ is different for each atom.

All photon-echo quantum memory protocols employ a procedure that rephases the atomic dipoles some time later, thereby recreating collective atomic coherence. In other words, the phases of all atoms become equal at some moment $t_e$. This triggers re-emission of the absorbed signal.
The initial distribution of spectral detunings $\delta_j$ allows for a classification of photon-echo quantum memory into two categories \cite{MT08}, which we describe now.

\subsubsection{Controlled reversible inhomogeneous broadening (CRIB)}
If the spectral distribution is continuous, the requirement of atom-independent phase evolution~$\int_0^{t_e}\delta_j(t) dt$ can only be achieved if, some time~$t'$ after absorption of the light, the resonance frequency of all atoms is actively changed from $\delta_{j1}$ to $\delta_{j2}$ so that $\delta_{j1} t'+\delta_{j2}(t_e- t')=\textrm{const}$ for all~$j$ (Fig.~4).

This approach to storage can be traced back to 1964, when the well-known spin echo \cite{Hahn1950} was extended to the optical domain \cite{Kurnit1964}. The two-pulse photon echo was then developed into time-variable storage of data pulses using three-pulse photon echoes \cite{Mossberg1979,EZM79,Mos82,Carlson1983}. However, conventional photon-echo does not allow efficient storage and recall of data encoded into few-photon pulses of light with high fidelity, due to an inherent amplification process \cite{RLSC08}. Yet, it recently inspired a quantum memory protocol that is now generally referred to as {\it controlled reversible inhomogeneous broadening} (CRIB) \cite{TAC+09}. Alternatively, the term {\it Gradient echo memory} (GEM) is used \cite{Hetet2008}. First proposed in 2001 for storage in atomic vapor \cite{MK01}, CRIB has meanwhile been adapted for solid-state storage of microwave photons \cite{MTH03}
and optical photons \cite{NK05,ALSM06,Kraus2006}.

The original proposal for CRIB is based on a hidden time-reversal symmetry in the Maxwell-Bloch equations that describe the evolution of the atom-light system during absorption and re-emission \cite{Kraus2006}. Reversing the evolution of the atom-light system requires that the detuning of all atoms is inverted, i.e. $\delta_{j2}=-\delta_{j1}$ (Fig.~4). Additionally, a mode-matching (or phase-matching) operation has to be performed, which consists of applying a phase shift $e^{-2ikz_j}$ to all atoms. This results in mapping the forward-traveling collective atomic coherence created during absorption of the forward-traveling light onto a backward-propagating coherence, which can lead to light emission in the backward direction. Another condition for perfect time reversal is that the optical depth of the medium must be sufficiently large to guarantee absorption of the incoming light.

If the conditions of atomic inversion, mode matching and large optical depth are not satisfied, symmetry arguments do not suffice to predict the memory performance. Of particular interest are the cases where the light is not completely absorbed, due to limited optical depth, and where the mode-matching operation is not implemented, resulting in the light being re-emitted in forward direction.
To find the performance, we distinguish between different types of inhomogeneous broadening. Two types have been analyzed. In \textit{transverse broadening}, the atomic absorption line is equally broadened for each position $z$. \textit{Longitudinal broadening} refers to the case where the atomic absorption line for each position $z$ is narrow, and the resonance frequency varies monotonically throughout the medium: $\delta_j=\chi z_j$.

Assuming transverse broadening and limited optical depth $\alpha L$, the efficiency for recall in the backward direction is given by \cite{MN04,SSAG07}
$\epsilon_b^{(t)}=\left(1-\exp\{-\alpha L\}\right)^2$.
For re-emission in the forward direction, is is \cite{SSAG07,LHLS08}
$\epsilon_f^{(t)}=(\alpha L)^2 \exp\{-\alpha L\}$.
In this case the maximum efficiency of $\epsilon_f^{(t)}=54\%$ is obtained for $\alpha L=2$. Note that the recalled pulse is time-reversed, resulting in an exchange of the leading and trailing end, regardless of the direction of recall.

Longitudinal broadening yields
$\epsilon_b^{(l)}=\epsilon_f^{(t)}=\left (1-\exp\{-(\alpha L)_{\rm eff}\}\right )^2$
where $(\alpha L)_{\rm eff}\propto\chi^{-1}$ characterizes the effective optical depth of the medium \cite{LHLS08,MS08,HLA+08}. It is interesting that the efficiency for forward recall can reach unity, despite the violation of time reversal: while the output pulse is a time-inverted image of the input signal \cite{Hetet2008}, it is re-emitted in forward direction. The recalled pulse features a frequency chirp, i.e.\ the fidelity of the retrieved mode deviates from unity \cite{MS08,HLA+08}.

Depending on the storage medium, the required change of detunings $\delta_j$ can be achieved in different ways. For atomic vapor where the inhomogeneous broadening is due to atomic motion, the atoms have to be forced to emit light in the backward direction using the above mentioned mode-matching operation, thereby inverting the Doppler shifts. In solids, the detunings of individual atoms depend on crystal defects and strain. Control over the detunings can be achieved by `tailoring' a narrow absorption line in the naturally broadened transition by optical pumping, followed by controlled and reversible broadening through position-dependent Stark or Zeeman shifts, as shown in Step 1 in Fig.~4.

In order to achieve a large delay-bandwidth product when working with a two-level system, we need, on the one hand, a narrow initial line, which determines the storage time (see sec.~\ref{sec:decoherence}). On the other hand, we need a large broadened line, which determines the bandwidth of the pulse to be stored. The required large artificial broadening compromises the optical depth, thereby impacting on the storage efficiency. This can be alleviated by working with broader initial lines and rapidly mapping optical coherence (between levels~$\ket b$ and~$\ket a$ in Fig.~1(a)) onto long-lived ground state coherence (between levels~$\ket b$ and~$\ket c$). This transfer can be accomplished using $\pi$ pulses, or a direct Raman transfer using additional, off-resonant control fields connecting levels $\ket a$ and~$\ket b$. If the control fields are counter-propagating, this procedure also implements the mode-matching operation forcing the retrieved signal to propagate backwards.

Photon-echo quantum memory with Raman transfer is sometimes referred to as {\it Raman echo quantum memory} (REQM) \cite{MT08,HHS+08,NRL+08,GB09,Hosseini09}. Beyond the possibility to work with relatively broad optical absorption lines, this approach may relax material requirements as the storage bandwidth depends not only on the controlled inhomogeneous broadening of the optical transition, but also on the Rabi frequency of the Raman control fields \cite{MT08}. Interestingly, reversible mapping of quantum states between light and atomic ensembles can be obtained for fields of arbitrary strengths, i.e.\ beyond the usual weak field, linear approximation \cite{MT08}.

CRIB was first demonstrated in 2006 using an Europium doped Y$_2$SiO$_5$ crystal, a reversible external electric field that generated longitudinal broadening, and macroscopic optical pulses recalled in the forward direction \cite{ALSM06}. As in all photon-echo based storage, the crystal was cooled to around 4K. Due to limited optical depth, the size of the recalled pulses was six orders of magnitude smaller than the one of the input pulses. Shortly after, the same group demonstrated that amplitude as well as phase information can be stored \cite{Alexander07}. The efficiency was similarly small. Since then, the memory performance has been improved tremendously, and a record efficiency of 66\% was recently reported using a Praseodymium-doped Y$_2$SiO$_5$ crystal and the configuration mentioned above \cite{HSLL09}.

The Europium and Praseodymium-doped crystals employed in these experiments feature a favorable level structure and radiative lifetimes for optical pumping, but have somewhat inconvenient transition frequencies around 580 and 606 nm, respectively. This requires working with frequency-stabilized dye lasers. Furthermore, the spectral width of the light to be stored is limited to a few MHz, due to small atomic level spacing in the ground and excited state multiplets. Recently, CRIB was implemented with telecommunication photons in an Erbium doped Y$_2$SiO$_5$ crystal \cite{LMdR+09} . This material features a more convenient transition at 1536 nm where standard diode lasers can be employed, but does not provide the same ease for tailoring the initial absorption line as Europium or Praseodymium doped Y$_2$SiO$_5$. The recall efficiency for weak coherent laser pulses was below 1\%.

CRIB based storage was recently combined with a direct Raman transfer  \cite{HHS+08,Hosseini09}. The experiments relied on macroscopic signal pulses and storage in Rubidium vapor. The first investigation established the feasibility and resulted in a recall efficiency of 1\%. In the second study, the efficiency could be increased up to 41\%. Exploiting the condition that pulse emission can only take place if all dipoles oscillate in phase and the Raman coupling beam is switched on, the authors also demonstrated that the order and the moments of recall of four stored pulses can be chosen at will. Hence, the approach could work as an optical random access memory for quantum information encoded into time-bin qubits.  Furthermore, beam splitting of input pulses was observed. The latter has also been demonstrated using EIT systems \cite{Appel2006,Vewinger2007,Campbell2009,NPG08}, traditional stimulated photon echoes \cite{Staudt06}, and photon-echo quantum memory based on atomic frequency combs \cite{dRAS+08}.

\begin{figure}
   \includegraphics[width=0.4\textwidth]{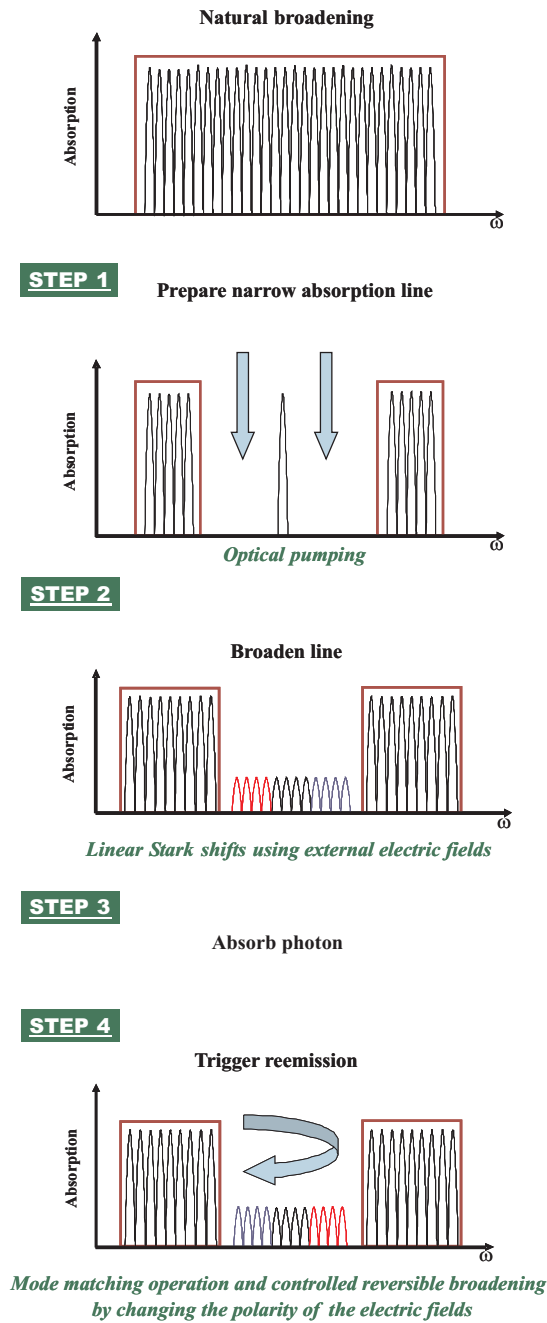}
\caption{CRIB-based quantum memory in solid state devices featuring optical centers with permanent electric dipole moments. In step 1, a narrow absorption line is created from an ensemble of absorbers with broad, inhomogeneously broadened absorption line. This is achieved by means of  an optical pumping (or spectral hole burning) procedure, which transfers population to auxiliary atomic levels. In step 2, the line is broadened in a controlled and reversible way using DC Stark shifts and position dependent external electric fields. This procedure goes along with a reduction of optical depth. Then, the signal to be stored is directed into the medium and absorbed (step 3). In step 4, re-emission is activated through the application of a mode (or phase) matching operation, i.e. a $2kz$ position dependent phase shift, and reversal of the detuning of each atom. This leads to backwards emission of the signal in a time-reversed version. (Figure from \cite{TAC+09})}
\label{figure CRIB}
\end{figure}

\subsubsection{Atomic frequency combs (AFC)}

In the protocol based on \textit{atomic frequency combs} (AFC),  the distribution of atoms over detuning $\delta$ is described by a periodic, comb-like structure with absorption lines spaced by multiples of $\Delta$ (see Fig.~5). Repetitive rephasing occurs at times $2\pi/\Delta$ when the phases accumulated by atomic dipoles in different ``teeth'' differ by multiples of $2\pi$. To inhibit re-emission after one fixed cycle time, and allow for long-term storage with on-demand readout, the excited optical coherence can be transferred temporarily to coherence between other atomic levels, e.g. ground state spin-levels, where the comb structure is not present. This condition is well satisfied in rare-earth-ion doped crystals.

The AFC approach originated from the discovery in the late seventies that photon echoes can be stimulated from accumulated frequency gratings \cite{HW79,RKS+83,Mitsunaga1991}.
The quantum memory protocol, proposed in 2008 \cite{ASdRG09}, is expected to enable time-variable storage of quantum states with unit efficiency and fidelity. Assuming transverse broadening, the efficiency of the AFC protocol can reach 100\% for recall in backward direction, and 54\% for recall in forward direction.

Compared to CRIB, AFC has the advantage of making better use of the available optical depth as fewer atoms need to be removed through optical pumping. Another advantage is the unlimited multi-mode capacity, provided that the natural broadened absorption line is sufficiently large. For instance, calculations suggest the possibility to store 100 temporal modes with an efficiency of 90\% in Europium Y$_2$SiO$_5$ \cite{ASdRG09}. For comparison, the multi-mode capacity in CRIB scales linearly with the absorption depth, and in EIT it is proportional to the square root thereof \cite{ASdRG09,NRL+08}.

AFC quantum memory with re-emission after 250 ns, pre-determined by the spacing in the frequency comb, was reported in 2008 \cite{dRAS+08}. The experiment relied on a Neodymium doped YVO$_4$ crystal, and recall in forward direction. The readout efficiency for weak coherent input states of ~$\sim$20 ns duration was 0.5\%, and a capacity of up to four temporal modes could be demonstrated. The post-selected storage fidelity for time-bin qubits in various states, defined by average state overlap discussed in sec.~\ref{sec:background}, exceeded 97\%.

A shortcoming in this experiment, namely the pre-determined emission timing, was overcome in 2009 \cite{Afzelius2009}. For on-demand recall, the initially excited optical coherence was temporally transferred to ground state coherence using two $\pi$ pulses with variable relative delay. 450 ns long macroscopic optical pulses could be stored in a Praseodymium doped Y$_2$SiO$_5$ crystal for up to 20 $\mu s$ with around 1\% efficiency.

In the same year, AFC-based storage was also demonstrated in a Thulium YAG crystal \cite{CAG09}. Better tailoring of the comb structure resulted in a storage efficiency of 9.1\%, i.e. an improvement by almost one order of magnitude.

\begin{figure}
 \includegraphics[width=.4\textwidth]{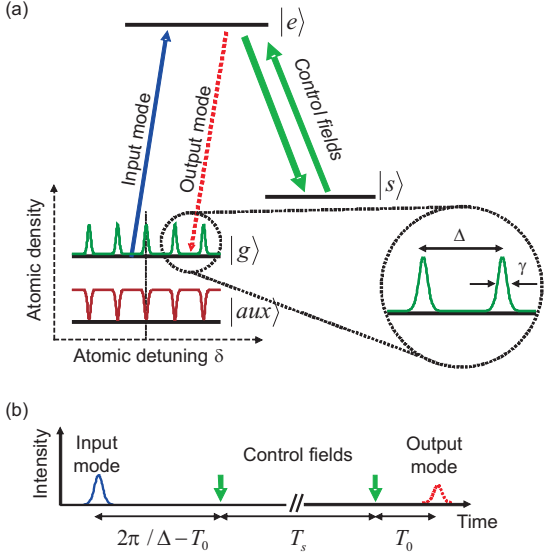}
 \caption{Quantum memory based on atomic frequency combs (AFC). (a) An inhomogeneously broadened absorption line (transition $\ket{g}-\ket{e}$) is tailored into an atomic frequency comb using frequency-selective optical pumping to the $\ket{aux}$ level. The peaks in the comb are characterized by width $\gamma$ and separation $\Delta$. (b) The collective dipole moment created by absorption of the input light rapidly dephases and, due to the discrete structure of the absorption profile tailored in (a), rephases after a time $2\pi/\Delta$. This results in the re-emission of the input light field. The application of a pair of counter-propagating control fields on the $\ket{e}-\ket{s}$ transition allows for long-time
storage, and efficient, on-demand recall in the backward direction after a storage time $T_s$. (Figure from \cite{ASdRG09})}
\label{figure_AFC}
\end{figure}

\subsection{Off-resonant Faraday interaction}

\begin{figure}
\includegraphics[width=\columnwidth]{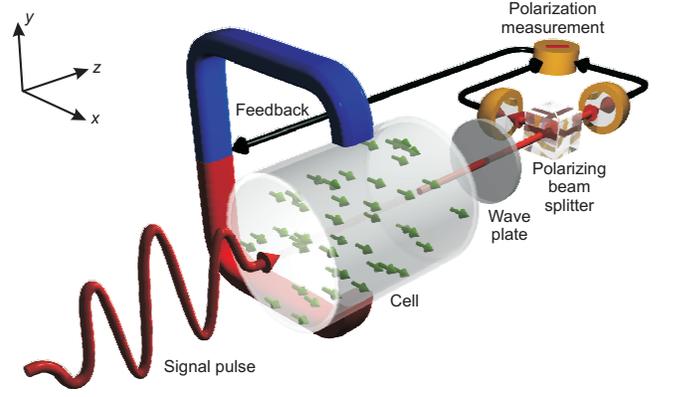}
 \caption{Quantum memory based on off-resonant Faraday interaction between light and atoms. The quantum information carried by the light is encoded in its polarization. After propagation through the cell, the field is subjected to a polarimetric measurement, whose result is then fed back to the atoms by applying a magnetic field pulse of a known magnitude and duration.
 }
 \label{PolzikFig}
\end{figure}

Consider an optical wave propagating through an ensemble of two-level atoms. When its detuning $\Delta$ from the atomic transition is sufficiently large ($\Delta\gg W_{\rm line}$), the real part of the susceptibility is inversely proportional to $\Delta$, whereas its imaginary part behaves as $\Delta^{-2}$ [Fig.~\ref{EITConcept}(b,c)]. Therefore an off-resonant wave will not excite any atoms, resulting in negligible  absorption, but may experience a significant phase shift. The quantum phase of the atoms will in turn be affected by the field.

This mutual effect of light and atoms can be utilized to construct an elegant implementation of quantum-optical memory. Consider a  signal wave with macroscopic linear polarization along the $y$ axis. The quantum information to be stored is encoded in the microscopic Stokes parameters $\hat S_2$ and $\hat S_3$ of the wave. The Stokes parameter $S_2$ is interpreted as the angle of polarization with respect to the $y$ axis, whereas $S_3$ is proportional to the collective spin of the photons. 

In order to implement storage, this field propagates, along the $z$ direction, through an off-resonant atomic gas. The atoms in the gas are initially prepared, by optical pumping, with the angular momentum $\hat{\vec J}$ oriented along the $x$ axis [Fig.~\ref{PolzikFig}(a)]. As a consequence of the angular momentum uncertainty relations, the projections $\hat J_y$ and $\hat J_z$ of the atomic collective are uncertain. In the following, we treat them as quantum operators.

The interaction between the light and atoms will lead to the following effects. First, the collective spin of the photons causes the atoms to rotate around the $z$ axis by a microscopic angle $\theta\propto\hat S_3$. We obtain
\begin{eqnarray}\label{light2atoms0}
\hat J_{y,{\rm out}}&=&\hat J_{y,{\rm in}}\cos\theta+\hat J_{x,{\rm in}}\sin\theta;\nonumber\\
\hat J_{x,{\rm out}}&=&\hat J_{x,{\rm in}}\cos\theta-\hat J_{y,{\rm in}}\sin\theta;\nonumber\\
\hat J_{z,{\rm out}}&=&\hat J_{z,{\rm in}}.
\end{eqnarray}
For a small $\theta$, we can assume $\cos\theta\approx 1$, $\sin\theta\approx\theta$. Furthermore, we can treat the macroscopic quantity $\hat J_x$ as a classical number and rewrite \eeqref{light2atoms0} as
\begin{eqnarray}\label{light2atoms}
\hat J_{y,{\rm out}}&=&\hat J_{y,{\rm in}}+\alpha \hat S_{3,{\rm in}};\nonumber\\
\hat J_{z,{\rm out}}&=&\hat J_{z,{\rm in}},
\end{eqnarray}
where $\alpha$ is some proportionality coefficient.

On the other hand, the optical wave's polarization will experience Faraday rotation due to the $z$ component of the angular momentum of the atoms:
\begin{eqnarray}\label{atoms2light}
\hat S_{2,{\rm out}}&=&\hat S_{2,{\rm in}}+\beta \hat J_{z,{\rm in}};\nonumber\\
\hat S_{3,{\rm out}}&=&\hat S_{3,{\rm in}}.
\end{eqnarray}
Here we again assumed that the rotation angle is small and treated the large $x$ polarization component as a classical number.

Equations \eqref{light2atoms} and \eqref{atoms2light} describe mutual interaction between the light and the atoms. We see that the information about the Stokes parameter $\hat S_3$ of the optical state is imprinted on the optical state. However, this does not yet constitute memory, because the atoms do not receive any information about $\hat S_3$; furthermore, the stored information is compromised by the uncertain values of $\hat J_y$ and $\hat J_z$.

In order to complete the memory protocol, we perform a polarimetric measurement of the pulse emerging from the atomic ensemble, thereby determining its Stokes parameter $S_{2,{\rm out}}$. Then we perform a feedback operation on the atoms, displacing its angular momentum $\hat J_z$ by the measured quantity as follows:
\begin{equation}\label{PolzikMemEq}
\hat J_z'=\hat J_{z,{\rm out}}-S_{2,{\rm out}}/\beta=-\hat S_{2,{\rm in}}/\beta.
\end{equation}
This displacement is performed, for example, by applying a magnetic field and causing the angular momentum to precess around the field direction.

Now both components of the optical polarization have been transferred to the atomic angular momentum. Although the $\hat J_y$ component is still ``contaminated'' by its initial noise $\hat J_{y,{\rm in}}$, this imperfection can be eliminated by initially preparing the atoms in the spin squeezed state \cite{KU93,KBM98}, so the uncertainty of $\hat J_{y,{\rm in}}$ is reduced.

Following initial theoretical papers in which off-resonant Faraday interaction between light and matter have been proposed as a tool for quantum information applications \cite{KP00,Duan00}, a theoretical proposal for quantum memory has been developed in 2003 by Kuzmich and Polzik \cite{KP03} and further elaborated in Ref. \cite{MHPC06}. In fact, these references propose a scheme in which the light passes through the atomic ensemble twice, in two different directions, which allows elimination of both measurement and feedback used in the scheme described above. 

Experimentally, the scheme with feedback has been implemented in 2004 by Julsgaard \iea \cite{JSF+04} using experimental tools developed earlier by the same group for the purpose of entangling two atomic ensembles \cite{JKP01}. The atoms were prepared without initial spin squeezing. However, the apparatus was shown to beat the classical benchmark for coherent states \cite{HPC05}, which was historically the first demonstration of quantum properties of an optical memory.

\subsection{Lifetime and decoherence}
\label{sec:decoherence}
A common feature in all approaches to quantum memory via atom-light interaction is storage of information in atomic coherence. All methods are thus prone to decoherence, which limits the storage time.

In the case of CRIB and AFC, the storage time is also affected by the width of the tailored absorption lines (see e.g. Step 1 in Fig.~4) as long as information is stored in coherence between the ground and the optically excited state \cite{SSAG07}. This width is fundamentally limited by the intrinsic homogeneous linewidth of the optical transition \cite{Macfarlane87,Macfarlane02,Sun02,Liu05,Macfarlane07,TAC+09}, and practically limited by laser line fluctuations and power broadening during optical pumping. Decoherence can be reduced by temporally mapping the optical coherence onto coherence between ground states, which are generally associated with smaller line-width and longer coherence time \cite{Afzelius2009}.

The storage time in Raman-type memory including EIT, DLCZ and REQM is limited by ground state decoherence. For instance, in vapor cells, the leading source of ground-state decoherence is the drift of atoms into and out of the laser beam. To reduce this effect, cells with inert buffer gases and/or paraffin-coated walls are generally used along with geometrically wider optical modes. In ultracold atoms confined in magneto-optical traps, decoherence often comes from the magnetic field. This field produces non-uniform Zeeman shift of atomic ground levels, leading to loss of the quantum phase in collective superpositions.

Significant improvement of the memory lifetime in atomic ensembles has recently been reported in two DLCZ experiments  \cite{BoZhao08,RZhao08}. The effect of the magnetic field has been minimized by using the atomic `clock' states as the ground states of the~$\Lambda$ system, i.e.\ such magnetic sublevels whose two-photon detuning is minimally affected by the Zeeman effect. Residual atomic motion has been eliminated in Ref.~ \cite{RZhao08} by transferring the atoms from a MOT into an optical lattice of a sufficiently small period. Ref.~ \cite{BoZhao08} instead used a collinear geometry for the four optical fields involved, in which case the dephasing due to atomic motion is greatly reduced. Storage times of 7 ms and 1 ms, respectively, have been reported in these experiments.

Very recently \cite{STT+09}, EIT-based storage of light has been demonstrated in an atomic Mott insulator --- a state of a collective of atoms filling a three-dimensional optical lattice, with one atom per lattice site. Absence of mechanical motion and uniformity of the magnetic field resulted in a storage lifetime of 238 ms, the current record for atomic media. The residual ground state coherence decay is likely due to heating in the optical lattice and atomic tunneling.

Even longer lived coherence can be achieved in rare-earth-ion doped crystals. Depending on the crystal and dopant, decoherence mechanisms vary. For instance, the dominant mechanism in Praseodymium-doped Y$_2$SiO$_5$ crystals is random Zeeman shifting of Pr$^{3+}$ ions due to fluctuating magnetic fields from Yttrium nuclei. Fortunately, this effect can be significantly reduced by operating in a uniform magnetic field of certain magnitude and direction. Using this approach, coherence times up to 82 ms have been reported \cite{FSL04}. This time was further increased to 30 s by adding dynamic decoherence control \cite{Fraval2005}.

\section{Conclusions}

Quantum memory for light constitutes a promising, rapidly developing research topic that builds on decades of research into atomic spectroscopy, quantum optics and material science. Recent results show efficient storage with high fidelity, long lifetime with on-demand recall, and high multi-mode capacities. Yet, these performance characteristics have to date been demonstrated using different storage media and protocols, and more effort is required for combining these benchmarks in a single setting. Once developed, such a device will be invaluable for quantum communication and cryptography as well as optical quantum computation.

\acknowledgments
This work was supported by iCORE, Quantum\textit{Works}, NSERC, CFI and GDC. A.I.L. is a
CIFAR Scholar and B.C.S. is a CIFAR Associate.


\begin{thebibliography}{1}
\bibitem {GRTZ02}
    Gisin, N., Ribordy, G., Tittel, W.\ \& Zbinden, H.
    Quantum cryptography.
    {\it Rev.\ Mod.\ Phys.} {\bf 74}, 145-195 (2002).
\bibitem {Mer07}
    Mermin, N.\ D.
    {\it Quantum Computer Science}
    (Cambridge University Press, Cambridge, 2007).
\bibitem{Lloyd96}
    Lloyd, S.
    Universal quantum simulators.
    {\it Science} {\bf 273}, 1073-1078 (1996).
\bibitem {KLM01}
    Knill, E., Laflamme, R.\ \& Milburn, G.\ J.
    A scheme for efficient quantum computation with linear optics.
    {\it Nature} {\bf 409}, 46-52 (2001).
\bibitem {LR02}
    Lund, A.\ P.\ \& Ralph, T.\ C.
    Nondeterministic gates for photonic single-rail quantum logic.
    {\it Phys.\ Rev.\ A} {\bf 66}, 032307 (2002).
\bibitem {BLS06}
    Berry, D.\ W., Lvovsky, A.\ I.\ \& Sanders, B.\ C.\
    Interconvertibility of single-rail optical qubits.
    {\it Opt.\ Lett.} {\bf 31}, 107-109 (2006).
\bibitem {MdRT+02}
    Marcikic, I. \iea
    Time-bin entangled qubits for quantum communication created by femtosecond pulses.
    {\it Phys.\ Rev.\ A} {\bf 66}, 062308 (2002).
\bibitem {Got06}
    Gottesman, D.
    Quantum Error Correction and Fault-Tolerance.
    in {\it Encyclopedia of Mathematical Physics} {\bf 4}, 196-201.
    Francoise, J.-P., Naber, G.\ L.\ \& Tsou, S.\ T., eds.
    (Elsevier, Oxford, 2006).
\bibitem {Sho95}
    Shor, P. W.
    Scheme for reducing decoherence in quantum computer memory,
    {\it Phys.\ Rev.\ A} {\bf 52}, R2493-R2496 (1995).

\bibitem{Lobino08}
    Lobino, M. \iea Complete characterization of quantum-optical processes. \textit{Science} \textbf{322}, 563-566 (2008).
\bibitem{Lobino09}
    Lobino, M., Kupchak, C., Figueroa E. \& Lvovsky A. I.,
    Memory for Light as a Quantum Process,
    \textit{Phys. Rev. Lett.} {\bf 102}, 203601 (2009).
\bibitem{HPC05}
    Hammerer, K., Wolf, M.M., Polzik, E.S., Cirac, J.I. Quantum benchmark for storage and transmission of coherent states. {\it Phys. Rev. Lett.} {\bf 94}, 150503 (2005)
\bibitem {HPJ+08}
    H\'{e}tet, G., Peng, A., Johnsson, M.\ T., Hope, J.\ J. \& Lam, P.\ K.
    Characterization of electromagnetically-induced-transparency-based
        continuous-variable quantum memories,
    {\it Phys.\ Rev.\ A} {\bf 77}, 012323 (2008).
\bibitem {ASdRG09}
    Afzelius, M., Simon, C., de Riedmatten, H. \& Gisin, N.
    Multimode quantum memory based on atomic frequency combs.
    {\it Phys.\ Rev.\ A} {\bf 79}, 052329 (2009).
\bibitem {NRL+08}
    Nunn, J. \iea
    Multimode Memories in Atomic Ensembles.
    {\it Phys.\ Rev.\ Lett.} \textbf{101}, 260502 (2008).
\bibitem{Raussendorf2001}
    Raussendorf, R., \& Briegel, H. J. A one-way quantum computer, {\it Phys. Rev. Lett.} {\bf86}, 5188-5191 (2001).
\bibitem{Kok2007}
    P. Kok, Munro, W. J., Nemoto, K., Dowling, J. P. \& Milburn, G. J., Linear optical quantum computing with photonic qubits.{\it Rev. Mod. Phys.} {\bf79}, 135-174 (2007).
\bibitem {BDCZ98}
    Briegel, H. J., D\"{u}r, W., Cirac, J.\ I. \& Zoller, P.
    Quantum repeaters: The role of imperfect local operations in quantum communication.
    {\it Phys.\ Rev.\ Lett.} {\bf 81}, 5932-5935 (1998).
\bibitem{Sangouard2009}
    Sangouard, M., Simon, C., de Riedmatten, H., \& Gisin, N. Quantum repeaters based on atomic ensembles and linear optics. arXiv:0906.2699.
\bibitem{Appel09}
    Appel, J. \iea Mesoscopic atomic entanglement for precision measurements beyond the standard quantum limit, {\it P Natl. Acad. Sci. USA} {\bf 106}, 10960-10965 (2009).
\bibitem{Hong1986}
    Hong, C.\ K.\ \& Mandel, L.
    Experimental realization of a localized one-photon state.
    \textit{Phys. Rev. Lett.} \textbf{56}, 58-60 (1986).
\bibitem {GRA86}
    Grangier, P., Roger, G.\ \& Aspect, A.
    Experimental Evidence for a Photon Anticorrelation Effect on a Beam Splitter:
        A New Light on Single-Photon Interferences.
    {\it Europhys.\ Lett.} {\bf 1}, 173-179 (1986).
\bibitem {LvHBZG07}
    Landry, O., van Houwelingen, J. A. W., Beveratos, A., Zbinden, H. \& Gisin, N.
    Quantum teleportation over the Swisscom telecommunication network.
    {\it J. Opt. Soc. Am. B} \textbf{24}, 398-403 (2007).
\bibitem {PF02a}
    Pittman, T.B. \& Franson, J.D.
    Single Photons on pseudodemand from stored parametric down-conversion.
    {\it Phys.\ Rev.\ A} \textbf{66}, 042303 (2002).
\bibitem {PF02b}
    Pittman, T.B. \& Franson, J.D.
    Cyclical quantum memory for photonic qubits.
    {\it Phys.\ Rev.\ A} \textbf{66}, 062302 (2002).
\bibitem {LR06}
    Leung, P.M. \& Ralph, T.C.
    Quantum memory scheme based on optical filters and cavities.
    {\it Phys.\ Rev.\ A} \textbf{74}, 022311 (2006).
\bibitem{MHN+97}
    Ma\^{i}tre, X. \iea
    Quantum Memory with a Single Photon in a Cavity.
    {\it Phys.\ Rev.\ Lett.} {\bf 79}, 769-772 (1997).

\bibitem{TNK+06}
    Tanabe, T., Notomi, M., Kuramochi, E., Shinya, A.\ \& Taniyama, H.
    Trapping and delaying photons for one nanosecond
    in an ultrasmall high-Q photonic-crystal nanocavity.
    {\it Nature Phot.} {\bf 1}, 49-52 (2006).
\bibitem{TNTK09}
    Tanabe, T., Notomi, M., Taniyama, H.\ \& Kuramochi, E.
    Dynamic Release of Trapped Light from an Ultrahigh-Q Nanocavity
    via Adiabatic Frequency Tuning.
    {\it Phys.\ Rev.\ Lett.} {\bf 102}, 043907 (2009).


\bibitem {JKLS02}
    A. Javan, O. Kocharovskaya, H. Lee \& M. O. Scully,
    Narrowing of EIT resonance in a Doppler Broadened Medium,
    {\it Phys.\ Rev.\ A} {\bf 66}, 013805 (2002).
\bibitem {BIH91}
	Boller, K.\ J., Imamoglu, A.\ \& Harris, S.\ E.
	Observation of electromagnetically induced transparency.
	{\it Phys.\ Rev.\ Lett.} {\bf 66}, 2593-2596 (1991).
\bibitem {KJYH95}
    Kasapi, A., Jain, M., Yin, G.\ Y.\ \& Harris, S.\ E.\,
    Electromagnetically Induced Transparency: Propagation Dynamics.
    {\it Phys.\ Rev.\ Lett.} {\bf 74}, 2447-2450 (1995).
\bibitem {FL02}
    Fleischhauer, M.\ \& Lukin, M.\ D.
    Quantum memory for photons: Dark-state polaritons.
    Phys.\ Rev.\ A {\bf 65}, 022314 (2002).
\bibitem {BKRY99}
    Budker, D., Kimball, D.\ F., Rochester, S.\ M. \& Yashchuk, V.\ V.
    Nonlinear Magneto-optics and Reduced Group Velocity of Light
        in Atomic Vapor with Slow Ground State Relaxation,
    {\it Phys.\ Rev.\ Lett.} {\bf 83}, 1767-1770 (1999).
\bibitem {HHDB99} Hau, L.\ V., Harris, S.\ E., Dutton, Z. \& Behroozi, C.\ H.
    Light speed reduction to 17 metres per second in an ultracold atomic gas.
    {\it Nature} {\bf 397}, 594-598 (1999)
\bibitem {BBB08}
	Burmeister, E.\ F., Blumenthal, D.\ J.\ \& Bowers, J.\ E.
	A comparison of optical buffering technologies.
	Optical Switching and Networking {\bf 5} 10-18 (2008)
\bibitem {FL00}
    Fleischhauer, M.\ \& Lukin, M.\ D.
    Dark-State Polaritons in Electromagnetically Induced Transparency.
    {\it Phys.\ Rev.\ Lett.} {\bf 84}, 5094-5097 (2000).
\bibitem {Luk03}
    Lukin, M.\ D.
    {\it Colloquium}: Trapping and manipulating photon states in atomic ensembles.
    Rev.\ Mod.\ Phys.\ {\bf 75}, 457-472 (2003).
\bibitem {FIM05}
    Fleischhauer, M.\, Imamoglu, A.\ \& Marangos, J.\ P.
    Electromagnetically induced transparency: Optics in Coherent Media.
    {\it Rev.\ Mod.\ Phys.} {\bf 77}, 633-673 (2005).
\bibitem {PFMW01} Phillips, D.\ F., A. Fleischhauer, A. Mair \& Walsworth, R.\ L.
    Storage of Light in Atomic Vapor.
    {\it Phys.\ Rev.\ Lett.} {\bf 86}, 783-786 (2001)
\bibitem {Hau01}
    Liu, C., Dutton, Z., Behroozi, C.\ H.\ \& Hau, L.\ V.
    Observation of coherent optical information storage in an atomic medium using halted light pulses.
    {\it Nature} {\bf 409}, 490-493 (2001).
\bibitem {GAF+07}
    Gorshkov, A.\ V., Andr\'e, A., Fleischhauer, M., S{\o}rensen, A.\ S.\ \& Lukin, M.\ D.
    Optimal storage of photon states in Optically Dense Atomic Media,
    {\it Phys.\ Rev.\ Lett.} {\bf 98}, 123601 (2007).
\bibitem {PNG08}
    Phillips, N.\ B.\, Novikova, I. \& Gorshkov, A.\ V.\,
    Optimal light storage in atomic vapor.
    {\it Phys.\ Rev.\ A} {\bf 78}, 023801 (2008).
\bibitem {NGP+07}
    Novikova, I. \iea
    Optimal Control of Light Pulse Storage and Retrieval
    {\it Phys.\ Rev.\ Lett.} {\bf 98}, 243602 (2007).
\bibitem {NPG08}
    Novikova, I.. Phillips, N.\ B.\ \& Gorshkov, A.\ V.\,
    Optimal light storage with full pulse-shape control.
    {\it Phys.\ Rev.\ A} {\bf 78}, 021802(R) (2008).
\bibitem {CVH09}
    Camacho, R.\ M., Vudyasetu, P.\ K.\ \& Howell, J.\ C.
    Four-wave-mixing stopped light in hot atomic rubidium vapour.
    {\it Nature Phot.} {\bf 3}, 103-106 (2009).
\bibitem{Ichimura98}
    Ichimura, K., Yamamoto, K.  and Gemma, N. Evidence for electromagnetically induced transparency in a solid medium, {\it Phys. Rev. A} {\bf 58}, 4116-4120 (1998).
\bibitem {TSS+02}
    Turukhin, A. V.  \iea
    {\it Phys.\ Rev.\ Lett.} {\bf 88}, 023602 (2002).
\bibitem {LFSM05}
    Longdell, J.\ J.\, Fraval, E., Sellars M.\ \& Manson, N.\ B.
    Stopped light with storage times greater than one second
        using electromagnetically induced transparency in a solid.
    {\it Phys.\ Rev.\ Lett.} {\bf 95}, 063601 (2005).
\bibitem {GG-NB+09}
    Goldner, Ph. \iea
    Long coherence lifetime and electromagnetically induced transparency
        in a highly-spin-concentrated solid.
    {\it Phys.\ Rev.\ A} {\bf 79}, 033809 (2009).
\bibitem {CMJ+06}
    Chaneli\`ere, T. \iea
    Storage and retrieval of single photons transmitted between remote quantum memories.
    {\it Nature} {\bf 438}, 833-836 (2006).
\bibitem {EAM+06}
    Eisaman, M.\ D. \iea
    Electromagnetically induced transparency with tunable single-photon pulses.
    {\it Nature} {\bf 438},837-841 (2006).
\bibitem {CDLK08}
    Choi, K. S., Deng, H., Laurat, J. \& Kimble,. H. J., Mapping photonic entanglement into and out of a quantum memory.
    {\it Nature} {\bf 452}, 67-71 (2008).
\bibitem{Kozuma04} Akamatsu, D., Akiba, K. and Kozuma, M. Electromagnetically Induced Transparency with Squeezed Vacuum, {\it Phys. Rev. Lett.} {\bf 92}, 203602 (2004).

\bibitem {HAA+08}
    Honda, K. \iea
    Storage and Retrieval of a Squeezed Vacuum
    {\it Phys.\ Rev.\ Lett.} {\bf 100}, 093601 (2008).
\bibitem{Honda09}
Arikawa, M. \iea M. Quantum memory of a squeezed vacuum for arbitrary frequency sidebands. arXiv:0905.2816
\bibitem {AFK+08}
    Appel, J., Figueroa, E., Korystov, D., Lobino, M. \& Lvovsky, A. I.,
    Quantum memory for squeezed light.
    {\it Phys.\ Rev.\ Lett.} {\bf 100}, 093602 (2008).
\bibitem {FVAL06}
    Figueroa, E., Vewinger, F., Appel, J. \& Lvovsky, A.\ I.\
    Decoherence of electromagnetically-induced transparency in atomic vapor.
    {\it Opt. Lett.} {\bf 31}, 2625-2627 (2006).
\bibitem {HHG+06}
    Hsu, M.\ T.\ L. \iea
    Quantum Study of Information Delay in Electromagnetically Induced Transparency.
    {\it Phys.\ Rev.\ Lett.} {\bf 97}, 183601 (2006).
\bibitem {PJB+05}
    Peng, A. \iea
    Squeezing and entanglement delay using slow light.
    {\it Phys.\ Rev.\ A} {\bf 71}, 033809 (2005)
\bibitem {PJB+05e}
    H\'{e}tet , G. \iea
    Erratum: Squeezing and entanglement delay using slow light [Phys. Rev. A 71, 033809 (2005)].
    Phys.\ Rev.\ A {\bf 74}, 059902(E) (2006).
\bibitem {FLK+09}
    Figueroa, E., Lobino, M., Korystov, D., Kupchak, C.\ \& Lvovsky, A. I.
    Propagation of squeezed vacuum under electromagnetically induced transparency,
    {\it New J. Phys.} {\bf 11}, 013044 (2009)
\bibitem {DLCZ01}
    Duan, L.-M., Lukin, M. D., Cirac, J.\ I. \& Zoller, P.
    Long-distance quantum communication with atomic ensembles and linear optics.
    {\it Nature} {\bf 414}, 413-418 (2001).


\bibitem {Kuzmich03} 	Kuzmich, A. A., Bowen, W. P., Boozer, A. D., Boca, A.,
		Chou, C. W., Duan, L.-M. \& Kimble, H. J.
	Generation of nonclassical photon pairs for
		scalable quantum communication with atomic ensembles
	{\it Nature} {\bf 423}, 731-734 (2003).

\bibitem {vanderWal03} 	van der Wal, C.\ H., Eisaman, M.\ D., Andr\'{e}, A., Walsworth, R.\ L., Phillips, D.\ F.,
		Zibrov, A.\ S.\ \& Lukin, M.\ D.
	Atomic Memory for Correlated Photon States.
	{\it Science} {\bf 301}, 196-200 (2003).

\bibitem {Jiang04} 	Jiang, W., Han, C., Xue, P., Duan, L.-M. \& Guo, G.-C.
	Nonclassical photon pairs generated from a room-temperature atomic ensemble.
	{\it Phys.\ Rev.\ A} {\bf 69}, 043819 (2004).

\bibitem {Chou04} Chou, C. W., Polyakov, S. V., Kuzmich, A. \& Kimble, H. J.
	Single-Photon Generation from Stored Excitation in an Atomic Ensemble.
	{\it Phys.\ Rev.\ Lett.} {\bf 92}, 213601 (2004).

\bibitem {Eisaman04} Eisaman, M. D., Childress, L., Andr\'{e}, A., Massou, F., Zibrov, A. S. \& Lukin, M. D.
	Shaping Quantum Pulses of Light Via Coherent Atomic Memory.
	{\it Phys.\ Rev.\ Lett.} {\bf 93}, 233602 (2004).

\bibitem {Polyakov04} Polyakov, S. V., Chou, C. W., Felinto, D. \& Kimble, H. J.
	Temporal Dynamics of Photon Pairs Generated by an Atomic Ensemble.
	{\it Phys.\ Rev.\ Lett.} {\bf 93}, 263601 (2004).

\bibitem {Laurat06} Laurat, J., de Riedmatten, J., Felinto, D., Chou, C.-W., Schomburg, E. W. \& Kimble, H. J.
	Efficient retrieval of a single excitation stored in an atomic ensemble.
	{\it Opt. Express} {\bf 14}, 6912-6918 (2006).

\bibitem {Balic05}  V. Bali\'{c}, V., Braje, D. A., Kolchin, P., Yin, G. Y. \& Harris, S.\ E.
	Generation of Paired Photons with Controllable Waveforms.
	{\it Phys.\ Rev.\ Lett.} {\bf 94}, 183601 (2005).

\bibitem {Braje04} D. A. Braje, D. A., Bali\'{c}, V., Goda, S., Yin G. Y. \&  Harris, S.\ E.
	Frequency Mixing Using Electromagnetically Induced Transparency in Cold Atoms.
	{\it Phys.\ Rev.\ Lett.} {\bf 93}, 183601 (2004).

\bibitem {Kimble09} Papp, S. B., Choi, K. S., Deng, H., Lougovski, P., van Enk, S. J. \& Kimble, H.J.
	Characterization of Multipartite Entanglement for One Photon Shared Among Four Optical Modes.
	{\it Science} {\bf 324}, 764-768 (2009).

\bibitem {Matsukevich06a} 	Matsukevich, D.\ N., Chaneli\'{e}re, T., Jenkins, S.\ D., Lan, S.\ Y., Kennedy, T.\ A.\ B. \& Kuzmich, A.
	Deterministic Single Photons via Conditional Quantum Evolution.
	{\it Phys.\ Rev.\ Lett.} {\bf 97}, 013601 (2006).

\bibitem {deRiedmatten06} de Riedmatten, H., Laurat, J., Chou, C.\ W., Schomburg, E.\ W., Felinto, D.\ \& Kimble, H.\ J.
	Direct Measurement of Decoherence for Entanglement
		between a Photon and Stored Atomic Excitation.
	{\it Phys.\ Rev.\ Lett.} {\bf 97}, 113603 (2006).

\bibitem {Chen06}Chen, S., Chen, Y.-A., Strassel, T., Yuan, Z.-S., Zhao, B., Schmiedmayer, J.\ \& Pan, J.-W.
	Deterministic and Storable Single-Photon Source Based on a Quantum Memory.
	{\it Phys.\ Rev.\ Lett.} {\bf 97}, 173004 (2006).

\bibitem {Vuletic05}  Black, A.\ T.,  Thompson, J.\ K.\ \& Vuleti\'{c}, V.
	On-Demand Superradiant Conversion of Atomic Spin Gratings
		into Single Photons with High Efficiency
	{\it Phys.\ Rev.\ Lett.} {\bf 95}, 133601 (2005).

\bibitem {Vuletic06} Thompson, J. K., Simon, J., Loh, H. \& Vuletic, V. A High-Brightness Source of
Narrowband, Identical-Photon Pairs. {\it Science} {\bf 313}, 74 (2006).

\bibitem {Vuletic07}  Simon, J., Tanji, H., Thompson, J.\ K.\  \& Vuleti\'{c}, V.
	Interfacing Collective Atomic Excitations and Single Photons.
	{\it Phys.\ Rev.\ Lett.} {\bf 98}, 183601 (2007).

\bibitem {Felinto06} D. Felinto \iea Conditional control of the quantum states of remote atomic memories for quantum networking. {\it Nature Phys.} {\bf 2}, 844 (2006).

\bibitem {Chaneliere07} Chaneli\`{e}re, T., Matsukevich, D.\ N., Jenkins, S.\ D., Lan, S.-Y., Zhao, R.,
		Kennedy, T.\ A.\ B.\ \& Kuzmich, A.
	Quantum Interference of Electromagnetic Fields from Remote Quantum Memories.
	{\it Phys.\ Rev.\ Lett.} {\bf 98}, 113602 (2007).

\bibitem {Yuan07} Yuan  Z.-S., Chen, Y.-A., Chen, S., Zhao, B., Koch, M., Strassel, T., Zhao, Y., Zhu, G.-J.,
		Schmiedmayer, J.\ \& Pan, J.-W.
	Synchronized Independent Narrow-Band Single Photons
		and Efficient Generation of Photonic Entanglement.
	{\it Phys.\ Rev.\ Lett.} {\bf 98}, 180503 (2007).



\bibitem {Matsukevich04} Matsukevich, D. N. \& Kuzmich, A.
	Quantum State Transfer Between Matter and Light.
	{\it Science} {\bf 306}, 663-666 (2004).
\bibitem {Kimblecomment} van Enk, S. \& Kimble, H. J.
	Comment on ``Quantum State Transfer Between Matter and Light''.
	{\it Science} {\bf 309}, 1187 (2005).
\bibitem {Kuzmichreply}    Matsukevich, D. N., \& Kuzmich, A. Response to Comment on ``Quantum State Transfer Between Matter and Light''. {\it Science} {\bf 309}, 1187 (2005).
\bibitem {Chou05} Chou, C. W. \iea Measurement-induced entanglement for excitation stored in remote atomic ensembles. {\it Nature} {\bf 438}, 828 (2005).

\bibitem{Chen2008}Chen, Y.-A. \iea Memory-build-in quantum teleportation with photonic and atomic qubits. \textit{Nature Phys.} \textbf{4}, 103-107 (2008).

\bibitem {Matsukevich05} Matsukevich, D. N. {et al.}. Entanglement of a Photon and a Collective Atomic Excitation. {\it Phys. Rev. Lett.} {\bf 95}, 040405 (2005).
\bibitem {Inoue06} Inoue, R., Kanai, N., Yonehara, T., Miyamoto, Y., Koashi, M.\ \& Kozuma, M.
	Entanglement of orbital angular momentum states
		between an ensemble of cold atoms and a photon
	{\it Phys.\ Rev.\ A} {\bf 74}, 053809 (2006).
\bibitem {Chen07} Chen, S.,  Chen, Y.-A., Zhao, B., Yuan, Z.-S., Schmiedmayer, J.\ \& Pan, J.-W.
	Demonstration of a Stable Atom-Photon Entanglement Source for Quantum Repeaters.
	{\it Phys.\ Rev.\ Lett.} {\bf 99}, 180505 (2007).

\bibitem {Lan07} S.-Y. Lan \iea Dual-Species Matter Qubit Entangled with Light. {\it Phys.\ Rev.\ Lett.} {\bf 98}, 123602 (2007).

\bibitem {Matsukevich06b} D. N. Matsukevich  \iea Entanglement of Remote Atomic Qubits. {\it Phys.\ Rev.\ Lett.} {\bf 96}, 030405 (2006).
\bibitem {Yuan08} Z.-S. Yuan \iea Experimental demonstration of a BDCZ quantum repeater node. {\it Nature} {\bf 454}, 1098 (2008).


\bibitem {Zhao07} Zhao, B., Chen, Z.-B., Chen, Y.-A., Schmiedmayer, J., \& Pan, J.-W.  {\it Phys.\ Rev.\ Lett.} {\bf 98}, 240502 (2007).

\bibitem {Chou07} Chou, C.-W. \iea Functional Quantum Nodes for Entanglement Distribution over Scalable Quantum Networks. {\it Science} {\bf 316}, 1316 (2007).


\bibitem {MT08}
    Moiseev, S.\ A. \& Tittel, W.
    Optical quantum memory with generalized time-reversible atom-light interactions, arXiv:0812.1730.
\bibitem {Hahn1950}
    Hahn, E.\ L.
    Spin Echoes.
    {\it Phys. Rev.} \textbf{80}, 580-594 (1950).
\bibitem {Kurnit1964}
    Kurnit, N., Abella, I.\ D. \& Hartmann, S.\ R.
    Observation of a Photon Echo.
    {\it Phys. Rev. Lett.} \textbf{13}, 567-568 (1964).
\bibitem {EZM79}
    Elyutin, S.O., Zakharov, S.M. \& Manykin, E.A.,
    Theory of formation of photon echo pulses.
    {\it Sov. Phys. JETP} \textbf{49}, 421-431 (1979).
\bibitem {Mos82}
    Mossberg, T.
    Time-domain frequency-selective optical data storage.
    {\it Opt. Lett.} \textbf{7}, 77-79 (1982).
\bibitem{Mossberg1979}Mossberg, T., Flusberg, A., Kachru, R., \& Hartmann, S. R. Total Scattering Cross Section for Na on He Measured by Stimulated Photon Echoes, {\it Phys. Rev. Lett.} {\bf 42}, 1665-1669 (1979).
\bibitem{Carlson1983}Carlson, L. J., Rothberg, L. J., Yodh, A. G., Babbitt, W. R. \& Mossberg, T. Storage and time reversal of light pulses using photon echoes. {\it Opt. Lett.} {\bf 8}, 483-485 (1983).
\bibitem {RLSC08}
    Ruggiero, J., Le Gou\"{e}t, J.-L, Simon, C.\ \& Chaneli\`{e}re, T.,
    Why the two-pulse photon echo is not a good quantum memory protocol.
    \textit{Phys. Rev. A} \textbf{79}, 053851 (2009).
\bibitem {TAC+09}
    Tittel, W. \iea
    Photon-Echo Quantum Memory in Solid State Systems.
    Laser \& Phot. Rev. DOI 10.1002/lpor.200810056
\bibitem{Hetet2008}
    H\'{e}tet, G., Longdell, J. J., Sellars, M. J., Lam, P. K., \& Buchler, B. C., Multimodal properties and dynamics of gradient echo memory. \textit{Phys. Rev. Lett.} \textbf{101}, 203601 (2008).
\bibitem {MK01}
    Moiseev, S.\ A. \& Kr\"oll, S.,
    Complete Reconstruction of the Quantum State of a Single Photon Wave Packet Absorbed by a Doppler-Broadened Transition
    {\it Phys.\ Rev.\ Lett.} \textbf{87}, 173601 (2001).
\bibitem {MTH03}
    Moiseev, S. A., Tarasov, V. F. \& Ham, B. S.,
    Quantum memory photon echo-like techniques in solids.
    {\it J. Opt. B: Quantum Semiclass. Opt.} \textbf{5}, S497-S502 (2003).
\bibitem {NK05}
    Nilsson, N. \& Kr\"oll, S.
    Solid state quantum memory using complete absorption and re-emission of photons by tailored and externally controlled inhomogeneous absorption profiles.
    {\it Opt. Comm.} \textbf{247}, 393-504 (2005).
\bibitem {ALSM06}
    Alexander, A.\ L., Longdell, J.\ J., Sellars, M.\ J. \& Manson, N.\ B.,
    Photon Echoes Produced by Switching Electric Fields.
    {\it Phys.\ Rev.\ Lett.} \textbf{96}, 043602 (2006).
\bibitem {Kraus2006}
    Kraus, B. \iea
    Quantum memory for nonstationary light fields based on controlled reversible inhomogeneous broadening
    {\it Phys.\ Rev.\ A} \textbf{73}, 020302 (2006).
\bibitem {MN04}
    Moiseev, S.\ A. \& Noskov, M.I.
    The possibilities of the quantum memory realization
        for short pulses of light in the photon echo technique.
    {\it Laser Phys. Lett.} \textbf{1}, 303-310 (2004).
\bibitem {SSAG07}
    Sanguard, N., Simon, C., Afzelius, M. \& Gisin, N.
    Analysis of a quantum memory for photons based on controlled reversible
        inhomogeneous.broadening.
    {\it Phys.\ Rev.\ A} \textbf{75}, 032327 (2007).
\bibitem {LHLS08}
    Longdell, J.\ J., H\'{e}tet, G., Lam, P.K. \& Sellars, M.\ J.
    Analytic treatment of controlled reversible inhomogeneous broadening
        quantum memories for light using two-level atoms.
    {\it Phys.\ Rev.\ A} {\bf 78}, 032337 (2008).
\bibitem {HLA+08}
    H\'{e}tet, M., Longdell, J.\ J., Alexander, A.\  L., Lam, P.\ K., Sellars, M.
    Electro-Optic Quantum Memory for Light Using Two-Level Atoms
    {\it Phys.\ Rev.\ Lett.} \textbf{100}, 023601 (2008).
\bibitem {MS08}
    Moiseev, S.\ A.\ \& Arslanov, N.M., Efficiency and fidelity of photon-echo quantum memory in an atomic system with longitudinal inhomogeneous broadening. Phys.\ Rev.\ A \textbf{78}, 023803 (2008).
\bibitem {HHS+08}
    H\'{e}tet, G. \iea
    Photon echoes generated by reversing magnetic field gradients in a rubidium vapor,
    {\it Opt. Lett.} \textbf{33}, 2323-2325 (2008).
\bibitem {GB09}
    Le Gou\"{e}t, J.-L. \& Berman, P.R.
    Raman scheme for adjustable bandwidth quantum memory.
    {\it Phys. Rev. A} {\bf 80}, 012320 (2009)
\bibitem{Hosseini09}
    Hosseini, M. \iea Coherent optical pulse sequencer for quantum applications. \textit{Nature} \textbf{461}, 241-245 (2009).
\bibitem{Alexander07}Alexander, A. L., Longdell, J. J. , Sellars, M. J. \& Manson, N. B. Coherent information storage with photon echoes produced by switching electric fields, \textit{J. Lumin.} \textbf{127}, 94-97 (2007).
\bibitem {HSLL09}
    Hedges, M.\ P., Sellars, M.\ J., Lee, Y.-M.\ \& Longdell, J.\ J.
    A Solid State Quantum Memory.
    Poster presentation at International Conference on Hole Burning, Single Molecule,
        and Related Spectroscopies: Science Applications (HBSM 2009), Palm Cove, Australia,
        22 to 27 June 2009.
\bibitem {LMdR+09}
    Lauritzen, B. \iea
    Solid state quantum memory for photons at telecommunication wavelengths,
    arXiv:0908.2348.
\bibitem{Appel2006}Appel, J., Marzlin, K.-P. \& Lvovsky, A. I. Raman adiabatic transfer of optical states in multilevel atoms. \textit{Phys. Rev. A} \textbf{73}, 013804 (2006).
\bibitem{Vewinger2007}Vewinger, F., Appel, J., Figueroa, E. \& Lvovsky, A. I. Adiabatic frequency conversion of quantum optical information in atomic vapor
    \textit{Opt. Lett.} \textbf{32}, 2771 - 2773 (2007).
\bibitem{Campbell2009} Campbell, G., Ordog, A. \& Lvovsky, A. I. Multimode electromagnetically-induced transparency on a single atomic line. New J. Phys. {\bf 11}, 103021 (2009)
\bibitem{Staudt06}
    Staudt, M. U. \iea Investigations of optical coherence properties in an erbium-doped silicate fiber for quantum state storage, \textit{Opt. Commun.} \textbf{266}, 720-726 (2006).
\bibitem {dRAS+08}
    de Riedmatten, H., Afzelius, M., Staudt, M.\ U., Simon, C.\ \& Gisin, N.,
    A solid-state light-matter interface at the single-photon level,
    {\it Nature} \textbf{456}, 773-777 (2008).
\bibitem {HW79}
    Hesselink, W.H. \& Wiersma, A.
    Picosecond Photon Echoes Stimulated from an Accumulated Grating.
    {\it Phys.\ Rev.\ Lett.} \textbf{43}, 1991-1994 (1979).
\bibitem {RKS+83}
    Rebane, A., Kaarli, R., Saari, P., Anijalk, A. \&Timpmann, K.
    Photochemical time-domain holography of weak picosecond pulses.
    {\it Opt. Comm.} \textbf{47}, 173-176 (1983).
\bibitem{Mitsunaga1991}
    Mitsunaga, M., Ryuzi, Y., \& Uesugi, N. Spectrally programmed stimulated photon echo. {\it Opt. Lett.} {\bf 16}, 264-266 (1991).
\bibitem{Afzelius2009}
    Afzelius, M. \iea Demonstration of atomic frequency comb memory for light with spin-wave storage. arXiv:0908.2309
\bibitem {CAG09}
    Chaneli\`{e}re, T., Afzelius \& M. \& Le Gou\"{e}t, J.-L.
    Efficient light storage in a crystal using an Atomic Frequency Comb,
    arXiv:0902.2048.
\bibitem{KU93}
    Kitagawa, M. \& Ueda, M.
    Squeezed spin states.
    {\it Phys.\ Rev.\ A} {\bf 47}, 5138-5143 (1993).
\bibitem{KBM98}
    Kuzmich, A., Bigelow, N.\ P.\ \& Mandel, L.\
    Atomic quantum non-demolition measurements and squeezing.
    {\it Europhys. Lett.} {\bf 42}, 481-486 (1998).
\bibitem {Duan00} Duan, L.-M., J. I. Cirac, P. Zoller, and E. S. Polzik, 2000,
Phys. Rev. Lett. 85, 5643.

\bibitem {KP00} Kuzmich, A., and E. Polzik, 2000, Phys. Rev. Lett. 85, 5639

\bibitem {KP03}
    Kuzmich, A. \& Polzik, E. S.
    Quantum Information with Continuous Variables (Kluwer, 2003), pp. 231--265.
\bibitem {MHPC06}
    Muschik, C. A., Hammerer, K., Polzik, E. S. \& Cirac, J.\ I.
    Efficient quantum memory and entanglement between light
        and an atomic ensemble using magnetic fields.
    {\it Phys.\ Rev.\ A} {\bf 73}, 062329 (2006).
\bibitem {JSF+04}
    Julsgaard, B., Sherson, J., Cirac, J.\ I., Fiur\'{a}\v{s}ek, J. \& Polzik, E. S.
    Experimental demonstration of quantum memory for light.
    {\it Nature} {\bf 432}, 482-486 (2004).
\bibitem {JKP01}
    Julsgaard, B., A. Kozhekin, A. \& E. S. Polzik.
    Experimental long-lived entanglement of two macroscopic objects.
    {\it Nature} {\bf 413}, 400-403 (2001).
\bibitem{Macfarlane87}Macfarlane, R. M. \& Shelby, R. M. in: Kaplyanskii, A.A. \& Macfarlane, R.M. (Eds.), Spectroscopy of Solids Containing Rare Earth Ions, North-Holland, Amsterdam, 1987.
\bibitem{Macfarlane02}Macfarlane, R. M.,     High-resolution laser spectroscopy of rare-earth doped insulators: a personal perspective, \textit{J. Lumin. } \textbf{100}, 1-20 (2002).
\bibitem{Sun02}Sun, Y., Thiel, C. W., Cone, R. L., Equall, R. W. \& Hutcheson, R. L. Recent progress in developing new rare earth materials for hole burning and coherent transient applications, \textit{J. Lumin. }\textbf{98}, 281-287 (2002).
\bibitem{Liu05}Spectroscopic Properties of Rare Earths in Optical Materials (Springer Series in Materials Science), Guokui Liu and Bernard Jacquier (Editors), Springer Berlin 2005.
\bibitem{Macfarlane07}Macfarlane, R. M. Optical Stark spectroscopy of solids, \textit{J. Lumin.} \textbf{125}, 156-174 (2007).
\bibitem {BoZhao08}
    Zhao, B. \iea A millisecond quantum memory for scalable quantum networks.
    {\it Nature Phys.} \textbf{5}, 95-99 (2008).
\bibitem {RZhao08}
    Zhao, R. \iea
    Long-lived quantum memory.
    {\it Nature Phys.} \textbf{5}, 100-104 (2008).
\bibitem {STT+09}
    Schnorrberger, U. \iea
    Electromagnetically Induced Transparency and Light Storage in an Atomic Mott Insulator,
    {\it Phys. Rev. Lett.} {\bf 103}, 033003 (2009)
\bibitem {FSL04}
    Fraval, E., Sellars, M.\ J.\ \& Longdell, J.\ J. Method of Extending Hyperfine Coherence Times in Pr$^{3+}$:Y$_2$SiO$_5$.
    {\it Phys.\ Rev.\ Lett.} {\bf 92}, 077601 (2004).
\bibitem{Fraval2005}Fraval, E., Sellars, M. J., \& Longdell, J.J. Dynamic Decoherence Control of a Solid-State Nuclear Quadrupole Qubit. \textit{Phys. Rev. Lett.} \textbf{95}, 030506 (2005).





\end{thebibliography}
\end{document}